\documentclass{PoS}
\usepackage{graphicx,amsmath,amssymb,wrapfig}

\title{3D structure of hadrons by generalized distribution amplitudes
and gravitational form factors}

\ShortTitle{3D Structure of Hadrons by GDAs}

\author{\speaker{S. Kumano}$^{\,\, a,b}$, Qin-Tao Song$^{\, a}$, 
        and O. V. Teryaev$^{\, c}$ \\
$^a$ KEK Theory Center, Institute of Particle and Nuclear Studies, KEK,\\
\ \ \            and Department of Particle and Nuclear Physics,
Graduate University for Advanced Studies \\
\ \ \  (SOKENDAI),       Ooho 1-1, Tsukuba, Ibaraki, 305-0801, Japan\\ 
$^b$ J-PARC Branch, KEK Theory Center,
     Institute of Particle and Nuclear Studies, KEK,\\
\ \ \ and Theory Group, Particle and Nuclear Physics Division, J-PARC Center,\\
\ \ \ 203-1, Shirakata, Tokai, Ibaraki, 319-1106, Japan\\
$^c$ Bogoliubov Laboratory of Theoretical Physics,
            Joint Institute for Nuclear Research,\\ 
\ \ \ 141980 Dubna, Russia}
\abstract{Generalized distribution amplitudes (GDAs) are one type
of three-dimensional structure functions, and they are related
to the generalized distribution functions (GPDs) by the $s$-$t$
crossing of the Mandelstam variables. The GDA studies 
provide information on three-dimensional tomography
of hadrons. The GDAs can be investigated by the two-photon process
$\gamma^* \gamma \to h\bar h$, and the GPDs are studied by
the deeply virtual Compton scattering $\gamma^* h \to \gamma h$.
The GDA studies had been pure theoretical topics, although
the GPDs have been experimentally investigated,
because there was no available experimental measurement.
Recently, the Belle collaboration reported their measurements on 
the $\gamma^* \gamma \to \pi^0 \pi^0$ differential cross section, 
so that it became possible to find the GDAs from their measurements. 
Here, we report our analysis of the Belle data 
for determining the pion GDAs.
From the GDAs, the timelike gravitational form factors
$\Theta_1 (s)$ and $\Theta_2 (s)$ can be calculated,
which are mechanical (pressure, shear force) and mass (energy) 
form factors, respectively. They are converted to the spacelike
form factors by using the dispersion relation, and then gravitational
radii are evaluated for the pion. 
The mass and mechanical radii are obtained from $\Theta_2$ and
$\Theta_1$ as
$\sqrt {\langle r^2 \rangle _{\text{mass}}} =0.56 \sim 0.69$ fm 
and $\sqrt {\langle r^2 \rangle _{\text{mech}}} =1.45 \sim 1.56$ fm,
whereas the experimental charge radius is
$\sqrt {\langle r^2 \rangle _{\text{charge}}} =0.672 \pm 0.008$ fm
for the charged pion. Future developments are expected in
this new field to explore gravitational physics 
in the quark and gluon level.
}

\FullConference{XXV International Workshop on Deep-Inelastic Scattering 
   and Related Subjects\\ 3-7 April 2017\\ University of Birmingham, UK}

\begin{document}

\section{Introduction}
\label{intro}
Understanding of three-dimensional (3D) structure functions for the nucleon
is one of hot topics in hadron physics. One of the major purposes 
of their studies is to understand the origin of nucleon spin 
including partonic orbital-angular-momentum contributions.
However, they could be used for other purposes such as
possible internal structure studies of exotic hadron candidates 
\cite{gdas-kk-2014}. Now, theoretical and experimental studies are 
focused on the generalized distribution functions (GPDs)
and transverse-momentum-dependent parton distributions (TMDs) 
as the 3D structure functions by deeply virtual Compton scattering
and semi-inclusive lepton scattering processes. There is another type of
3D structure functions called generalized distribution amplitudes (GDAs)
which can be measured by timelike processes, typically by
the two-photon process $\gamma^* \gamma \to h\bar h$ to produce a hadron pair
in the final state \cite{gpds-gdas}.
It is the $s$-$t$ crossed channel to the virtual Compton scattering
$\gamma^* h \to \gamma h$. There was no experimental measurement
for studying the GDAs until recently, so that they had been 
investigated as a theoretical subject.
However, time has come to find the GDAs for hadrons because
the Belle collaboration reported measurements on
the two-photon process $\gamma^* \gamma \to \pi^0 \pi^0$
recently \cite{Masuda:2015yoh}. It became possible
to extract appropriate GDAs from experimental measurements.
Once the GDAs are obtained, it is possible to calculate
gravitational form factors for hadrons.

Here, we determine quark GDAs and gravitational form factors
for $\pi^0$ from an analysis of the Belle data in the leading order 
of the QCD running coupling constant and 
the leading twist \cite{gdas-kst-2017}. This is the first attempt 
to extract the GDAs and the gravitational form factors
from actual measurements. In Sec.\,\ref{formalism}, we explain
our formalism to describe the $\gamma^* \gamma \to \pi^0 \pi^0$
cross section in terms of the GDAs.
In Sec.\,\ref{determination}, our analysis results are shown
for the GDAs, and then the gravitational form factors and radii
are calculated from the determined GDAs.
Our studies are summarized in Sec.\,\ref{summary}.

\section{Theoretical formalism}
\label{formalism}

\subsection{GDAs and gravitational form factors}

\begin{figure}[b!]
\vspace{-0.30cm}
\begin{minipage}{\textwidth}
\begin{tabular}{lcl}
\hspace{1.25cm}
\begin{minipage}[c]{0.45\textwidth}
    \includegraphics[width=6.0cm]{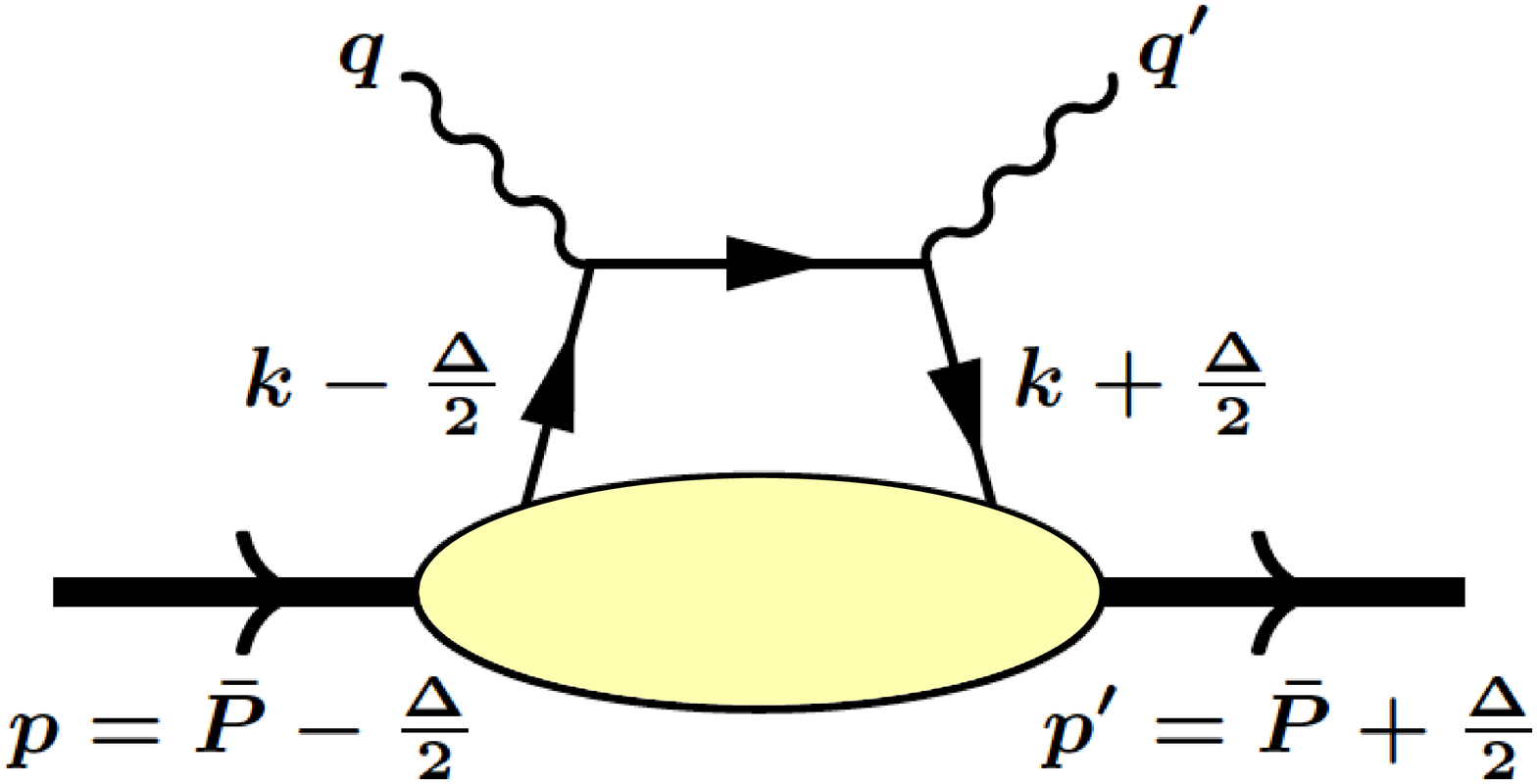}
\end{minipage} 
\hspace{0.00cm}
\begin{minipage}[c]{0.45\textwidth}
    \includegraphics[width=5.0cm]{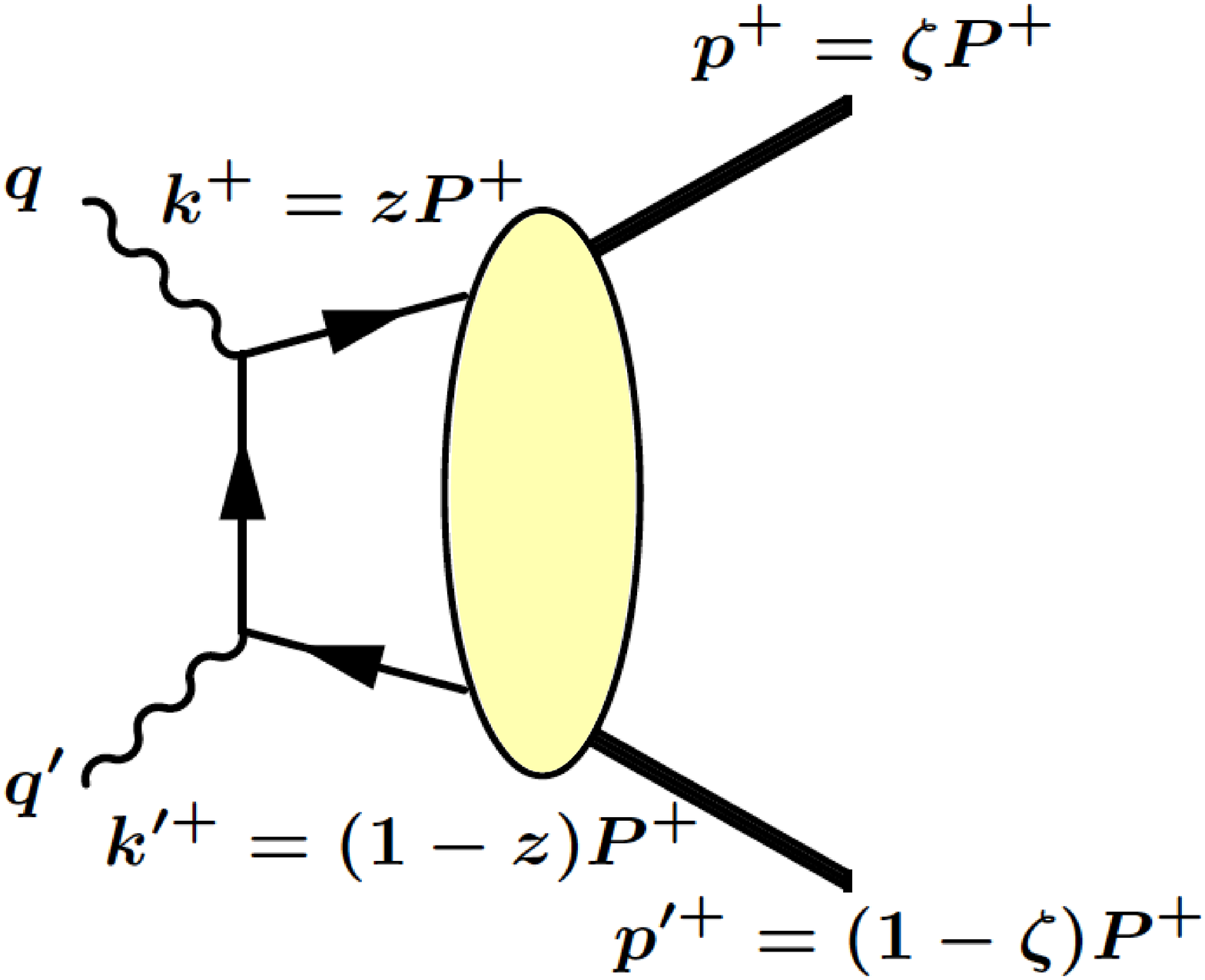}
\end{minipage}
\end{tabular}
\end{minipage}
\ \vspace{-0.00cm}\hspace{4.2cm}
$(a)$ \hspace{5.2cm} $(b)$
\vspace{-0.2cm}
\caption{$(a)$ Virtual Compton process to probe GPDs \ \ 
$(b)$ Two-photon process to probe GDAs.}
\label{fig:gpd-fig}
\vspace{-0.0cm}
\end{figure}

The GPDs can be measured by the deeply virtual Compton scattering
(DVCS) shown in Fig.1$(a)$, whereas the the GDAs can be investigated
by the two-photon process in Fig.1$(b)$. They are related
by the $s$-$t$ crossing of the Mandelstam variables.
The pion, photon, and quark momenta are denoted in Fig.\,\ref{fig:gpd-fig}.
The processes are factorized into the hard perturbative QCD part
and the soft one described by the GPDs and GDAs 
if the conditions,
$Q^2 \gg |t|, \ \Lambda_{\text{QCD}}^2$ for the DVCS and
$Q^2 \gg W^2, \ \Lambda_{\text{QCD}}^2$ for the two photon process,
are satisfied.
The variables $t$ and $W^2$ are defined by
$t=(p-p')^2$ and $W^2 =(p+p')^2 =s$. 
The pion GPDs, for example for $\pi^0$,
are defined by off-forward matrix elements
of quark and gluon operators with a lightcone separation as
\begin{align}
 \int\frac{d y^-}{4\pi} \, e^{i x \bar P^+ y^-}
 \left< \pi^0 (p') \left| 
\overline{q}(-y/2) \gamma^+ q(y/2) 
 \right| \pi^0 (p) \right> \Big |_{y^+ = \vec y_\perp =0}
 =  H_q^\pi (x,\xi,t) .
\label{eqn:gpd-pi}
\end{align}
Here, $x$ is given by $x = Q^2 /(2p \cdot q)$ with $Q^2=-q^2$,
and $\xi$ is the skewdness parameter given by
$\xi = \bar Q^2 /(2\bar P \cdot \bar q)$
with $\bar q=(q+q')/2$, $\bar P=(p+p')/2$, and $\bar Q^2 = - \bar q^2$.
The link operator for the color gauge invariance 
is not explicitly written in Eq.\,(\ref{eqn:gpd-pi})
just for simplicity.
The pion is not a stable target, so that the usual DVCS process
cannot be used for measuring its GPDs. Nevertheless, we show the definition
in order to compare with the pion GDAs, which are investigated
in this work.
The quark GDAs are defined by the matrix element 
of the same operator in defining the GPDs 
between the vacuum and the hadron pair:
\begin{align}
& \Phi_q^{\pi^0 \pi^0} (z,\zeta,W^2) 
= \int \frac{d y^-}{2\pi}\, e^{i (2z-1)\, P^+ y^- /2}
  \langle \, \pi^0 (p) \, \pi^0 (p') \, | \, 
 \overline{q}(-y/2) \gamma^+ q(y/2) 
  \, | \, 0 \, \rangle \Big |_{y^+=\vec y_\perp =0} \, .
\label{eqn:gda-def}
\end{align}
Here, the variable $z$ is defined by
$z = k \cdot q'/(P \cdot q')$ with the total momentum $P=p+p'$,
and $\zeta$ is given by $\zeta = p \cdot q'/(P \cdot q') = (1+\beta \cos \theta)/2$
with the pion velocity 
$\beta =|\vec p \,|/p^0 = \sqrt{1-4m_\pi^2/W^2}$
and the scattering angle $\theta$ in the center-of-mass frame
of final pions.

The GPDs and GDAs contain information on form factors
of not only electromagnetic interactions but also
gravitational interactions.
It is understood by taking $n$-th moments of the GPDs and GDAs.
For example, they are given for the GDAs as
\begin{align}
\! \!
2(P^+/2)^{n} \! \! \int_0^1 dz \, (2z-1)^{n-1} 
  \! \!  \int\frac{d y^-}{2\pi}e^{i (2z-1) P^+ y^- /2}
\overline{q}(-y/2) \gamma^+ q(y/2) \Big |_{y^+ = \vec y_\perp =0}
\! \! \!
 = \overline q (0) \gamma^+ \!
 \left ( i \overleftrightarrow \partial^+  \right )^{n-1} 
\! \! \!
 q(0) ,
\label{eqn:tensor-int}
\end{align} 
\vspace{-0.7cm}
\begin{wrapfigure}[10]{r}{0.54\textwidth}
   \vspace{-0.6cm}
   \begin{center}
     \includegraphics[width=7.0cm]{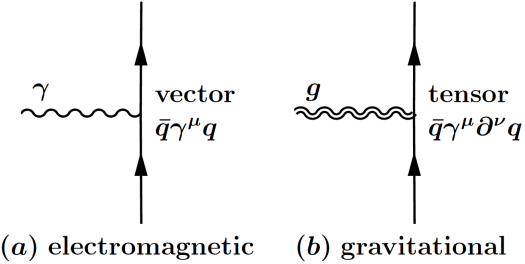}
   \end{center}
\vspace{-0.60cm}
\caption{Electromagnetic and gravitational form factors.}
\label{fig:electro-grav}
\vspace{-0.7cm}
\end{wrapfigure}
where the derivative $\overleftrightarrow \partial$ is defined by
$f_1 \overleftrightarrow \partial f_2 
 = [ f_1 (\partial f_2)  - (\partial f_1) f_2 ]/2$.
Equation (\ref{eqn:tensor-int}) indicates that 
the operator is the vector-type electromagnetic current 
and the energy-momentum tensor of a quark for $n=1$ and $n=2$, 
respectively, as shown in Fig.\,\ref{fig:electro-grav}.
Therefore, if the GPDs and/or GDAs are determined,
the gravitational form factors are obtained from them.

In fact, the second moments of the $\pi^0$ GDAs are given by
\begin{align}
\int_0^1 dz (2z -1) \, 
\Phi_q^{\pi^0 \pi^0} (z,\,\zeta,\,W^2) 
 = \frac{2}{(P^+)^2} \langle \, \pi^0 (p) \, \pi^0 (p') \, | \, T_q^{++} (0) \,
       | \, 0 \, \rangle ,
\label{eqn:integral-over-z}
\end{align}
where the quark energy-momentum tensor is
$ T_q^{\,\mu\nu} (x) = \overline q (x) \, \gamma^{\,(\,\mu} 
   i \overleftrightarrow D^{\nu)} \, q (x)$. 
Here, $D^\mu$ is the covariant derivative 
$D^\mu = \partial^{\,\mu} -ig \lambda^a A^{a,\mu}/2$ defined by 
the QCD coupling constant $g$ 
and the SU(3) Gell-Mann matrix $\lambda^a$.
Using the momenta $P=p+p'$ and $\Delta=p'-p$, 
we express the matrix element of the energy momentum tensor
in terms of the gravitational form factors $\Theta_1$ and $\Theta_2$
as
\begin{align}
\langle \, \pi^0 (p) \, \pi^0 (p') \, | \, T_q^{\mu\nu} (0) \, | \, 0 \, \rangle 
= \frac{1}{2} 
  \left [ \, \left ( s \, g^{\mu\nu} -P^\mu P^\nu \right ) \, \Theta_{1, q} (s)
                + \Delta^\mu \Delta^\nu \,  \Theta_{2, q} (s) \,
  \right ] .
\label{eqn:emt-ffs-timelike-0}
\end{align}
Therefore, if the GDAs (or GPDs) are determined, we can obtain
the gravitational form factors of the pion and subsequently
its gravitational radii. It is interesting to investigate 
the gravitational radii of hadrons in comparison with
charge radii because their physics origins are different.

\subsection{$\gamma^* \gamma \to \pi^0 \pi^0$ cross section in terms of GDAs}

The cross section for the pion-pair production process 
$\gamma^* (q) \gamma (q') \rightarrow \pi^0 (p) \pi^0 (p')$ is
expressed by the matrix element ${\cal M}$ as 
\vspace{-0.2cm}
\begin{align}
d\sigma = \frac{1}{4 q\cdot q'}
\underset{\lambda, \lambda'}{\overline\sum}
| {\cal M} (\gamma^* \gamma \to \pi^0 \pi^0 ) |^2 \,
\frac{d^3 p}{(2\pi)^3 \, 2E_p}  \frac{d^3 p'}{(2\pi)^3 \, 2E_{p'}} 
(2\pi)^4 \delta^4(q+q'-p-p') .
\label{eqn:cross-section}
\\[-0.80cm] \nonumber
\end{align}
Here, one of the initial photons is taken on mass shell (${q'}^2=0$).
The matrix element ${\cal M} (\gamma^* \gamma \to \pi^0 \pi^0)$
is given by the hadron tensor ${\cal T}_{\mu\nu}$ 
and the photon polarization vector $\epsilon^\mu(\lambda)$ as
$ i {\cal M} (\gamma^* \gamma \to \pi^0 \pi^0 ) 
   = \epsilon^\mu(\lambda) \, \epsilon^\nu(\lambda') \, {\cal T}_{\mu\nu} $,
where ${\cal T}_{\mu \nu }$ is expressed by 
the electromagnetic current $J_\mu ^{em}(y)$,
and then it is given by the  quark GDAs as
\vspace{-0.1cm}
\begin{align}
{\cal T}_{\mu \nu } 
& = i \! \! \int \! d^4 y \, {e^{ - iq \cdot y}} \!
\left\langle \pi^0 (p) \pi^0 (p') \! \left| 
{TJ_\mu ^{em}(y)J_\nu ^{em}(0)} \right|0 \right\rangle 
\nonumber \\[-0.10cm]
& =  - g_{T}^{\, \mu \nu}{e^2} 
\sum\limits_q \frac{{e_q^2}}{2} \!
\int_0^1 \! \! {dz} \frac{{2z - 1}}{{z(1 - z)}}
\Phi_q^{\pi^0 \pi^0}(z,\zeta ,{W^2}) . 
\label{eqn:matrix}
\\[-0.80cm] \nonumber
\end{align}
in the leading order and leading twist with 
$ g_{T}^{\, \mu \nu} = -1 $  for $\mu=\nu=1, \ 2$ and
$ g_{T}^{\, \mu \nu} = 0 $  for $\mu$, $\nu=\,$others.
The quark GPDs are expressed as
\begin{align}
\Phi_q^{\pi ^0 \pi^0} (z, \zeta, W^2) & = 
           N_\alpha z^\alpha(1-z)^\alpha (2z-1) \,
 [\widetilde B_{10}(W^2) + \widetilde B_{12}(W^2) P_2(\cos \theta)] , 
\label{eqn:gda-parametrization}
\\[-0.60cm] \nonumber
\end{align}
where S- and D-wave terms are expressed as
$\widetilde B_{10}(W^2)$ and $\widetilde B_{12}(W^2)$,
and $P_2(\cos \theta)$ is the Legendre polynomial.
The functions $\widetilde B_{nl}(W^2)$ are described by the continuum term
and resonance ones: 
$\widetilde B_{10}(W^2)= \text{continuum} + \text{resonance ($f_0$)}$,
$\widetilde B_{12}(W^2)= \text{continuum} + \text{resonance ($f_2$)}$.
We try to fix the resonance terms as much as possible 
from other theoretical and experimental studies.
The GPDs are expressed by a number of parameters, which are determined
by a $\chi^2$ analysis of the Belle experimental data for
$\gamma^* \gamma \to \pi^0 \pi^0$.
Because of the page limitation, we do not explain the details
of the GDA parametrization. One of the parameters is the 
cutoff $\Lambda$ in the overall form factor of the continuum part
$ F^{\,\pi}_q (W^2) = 1 / [ 1 + (W^2-4 m_\pi^2)/\Lambda^2 ]^{n-1}$
with the constituent-counting factor $n=2$ \cite{counting}.
The details should be found in Ref.\,\cite{gdas-kst-2017}.

\section{Determination of GDAs and gravitational form factors for pion}
\label{determination}

\begin{figure}[h!]
\vspace{0.10cm}
\begin{minipage}{\textwidth}
\begin{tabular}{lc}
\hspace{-0.30cm}
\begin{minipage}[c]{0.48\textwidth}
   \vspace{-0.2cm}
   \begin{center}
     \includegraphics[width=7.2cm]{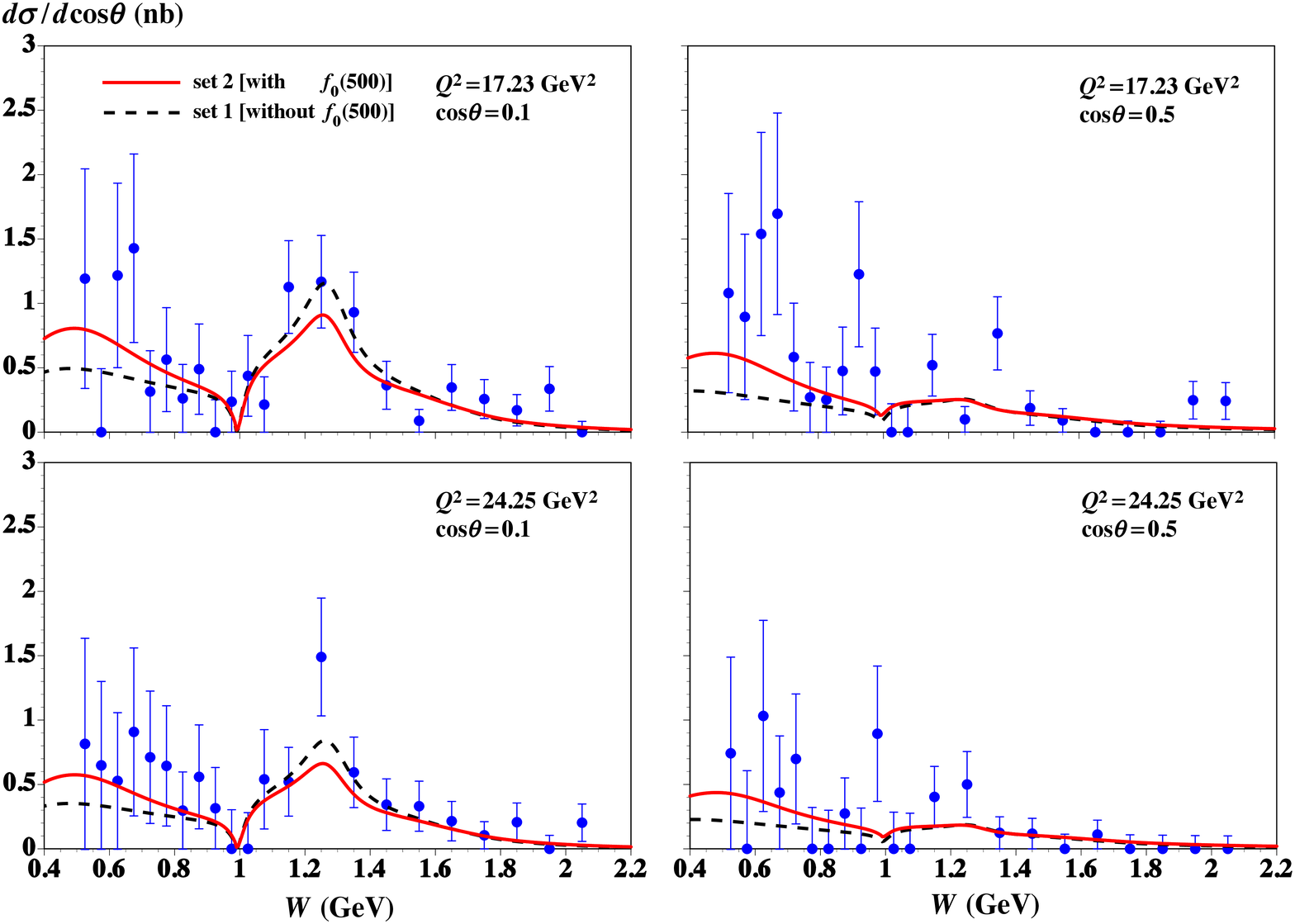}
   \end{center}
\vspace{-0.80cm}
\caption{Comparison with Belle data \cite{gdas-kst-2017}.}
\label{fig:belle-comparison}
\vspace{-0.4cm}
\end{minipage} 
\hspace{-0.0cm}
\begin{minipage}[c]{0.51\textwidth}
    \vspace{-0.2cm}
   \begin{center}
    \includegraphics[width=6.3cm]{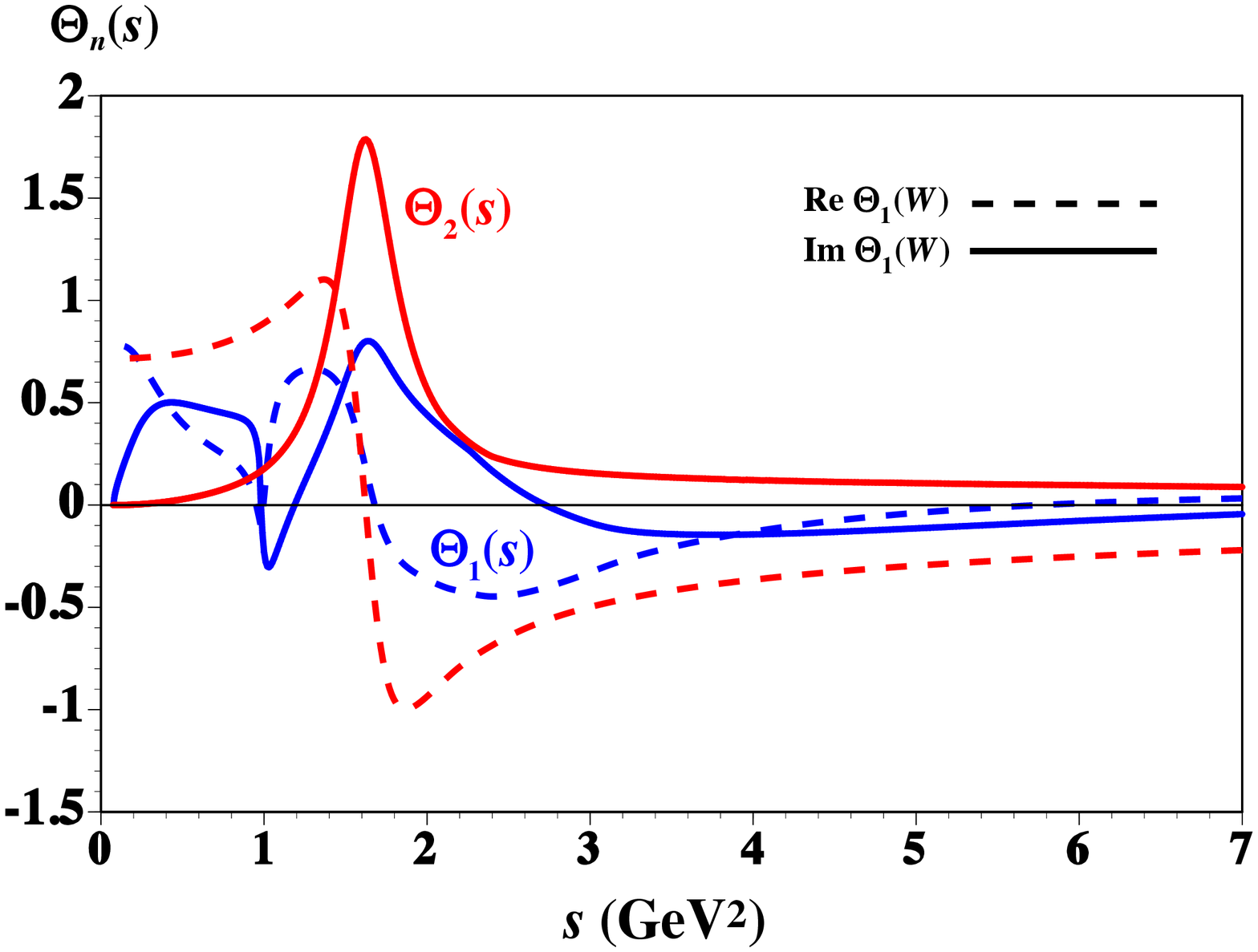}
   \end{center}
\vspace{-0.80cm}
\caption{Timelike gravitational form factors for $\pi$ \cite{gdas-kst-2017}.}
\label{fig:theta12reim-s}
\vspace{-0.4cm}   
\end{minipage}
\end{tabular}
\vspace{0.20cm}
\end{minipage}
\end{figure}

Our analysis results are compared with the Belle data on
$\gamma^* \gamma \rightarrow \pi^0 \pi^0$ in 
Fig.\,\ref{fig:belle-comparison}.
First, the $f_0 (980)$ resonance is not included in our analysis.
The decay constant of $f_0 (980)$ has been calculated theoretically
by assuming that it is a $q\bar q$ state.
If such $q\bar q$-type decay constant is
used, it is much larger than the Belle data at $W \simeq 1$ GeV.
It means that $f_0 (980)$ should not be interpreted by 
the $q\bar q$ configuration and it is possibly
a tetraquark hadron or a $K\bar K$ molecule
as other studies indicate \cite{f0-4q}.
In Fig.\,\ref{fig:belle-comparison}, the dotted curves indicate
our cross sections without $f_0 (500)$, and the solid
ones are obtained by including a $f_0 (500)$ contribution.
It is difficult to interpret the data without $f_0 (500)$
at small $W$ ($<0.8$ GeV). The overall agreement is obtained
with the experimental measurements.

\begin{figure}[b!]
\vspace{-0.30cm}
\begin{minipage}{\textwidth}
\begin{tabular}{lcl}
\hspace{0.70cm}
\begin{minipage}[c]{0.45\textwidth}
    \includegraphics[width=6.0cm]{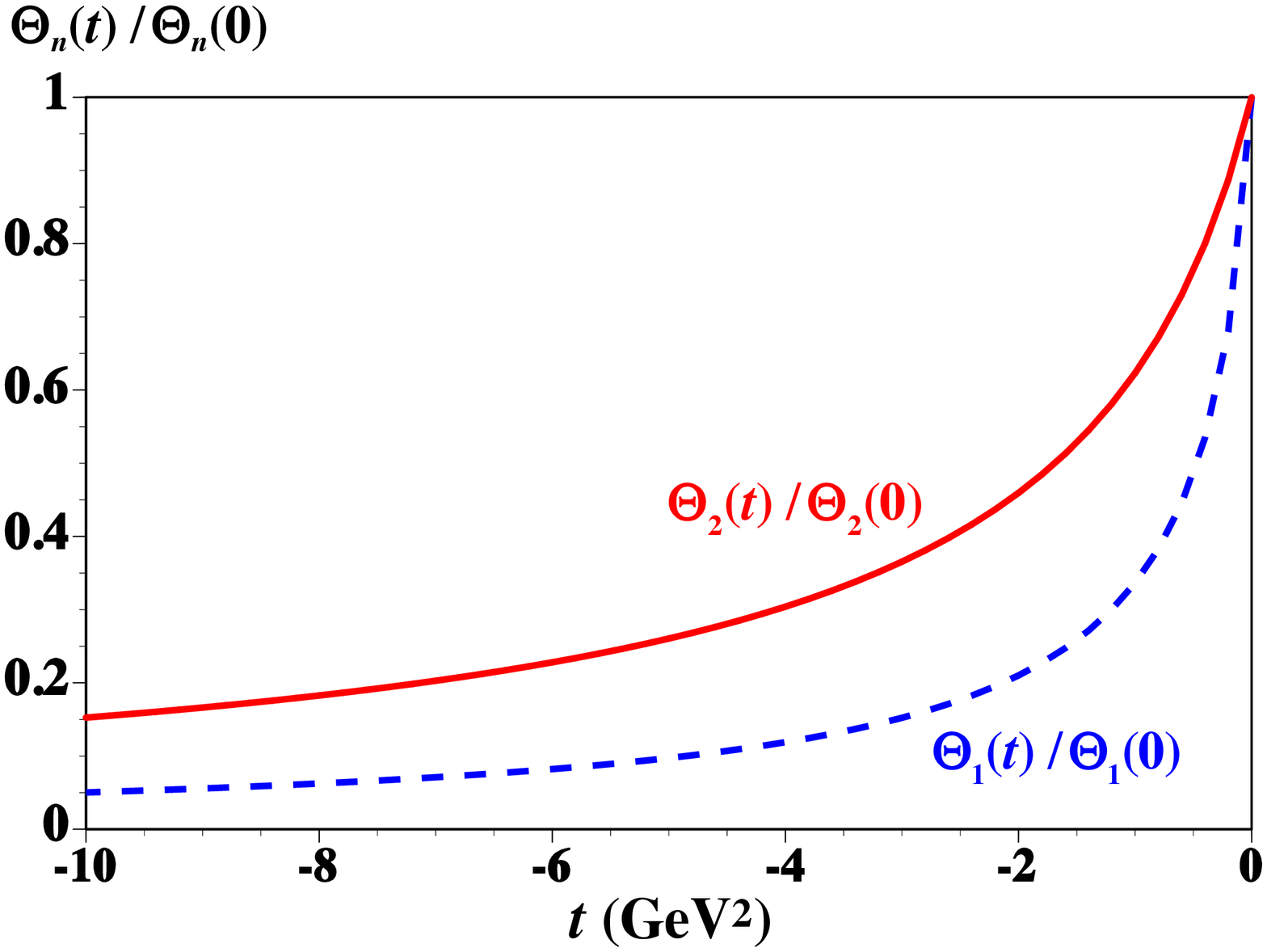}
\end{minipage} 
\hspace{0.00cm}
\begin{minipage}[c]{0.45\textwidth}
    \includegraphics[width=6.0cm]{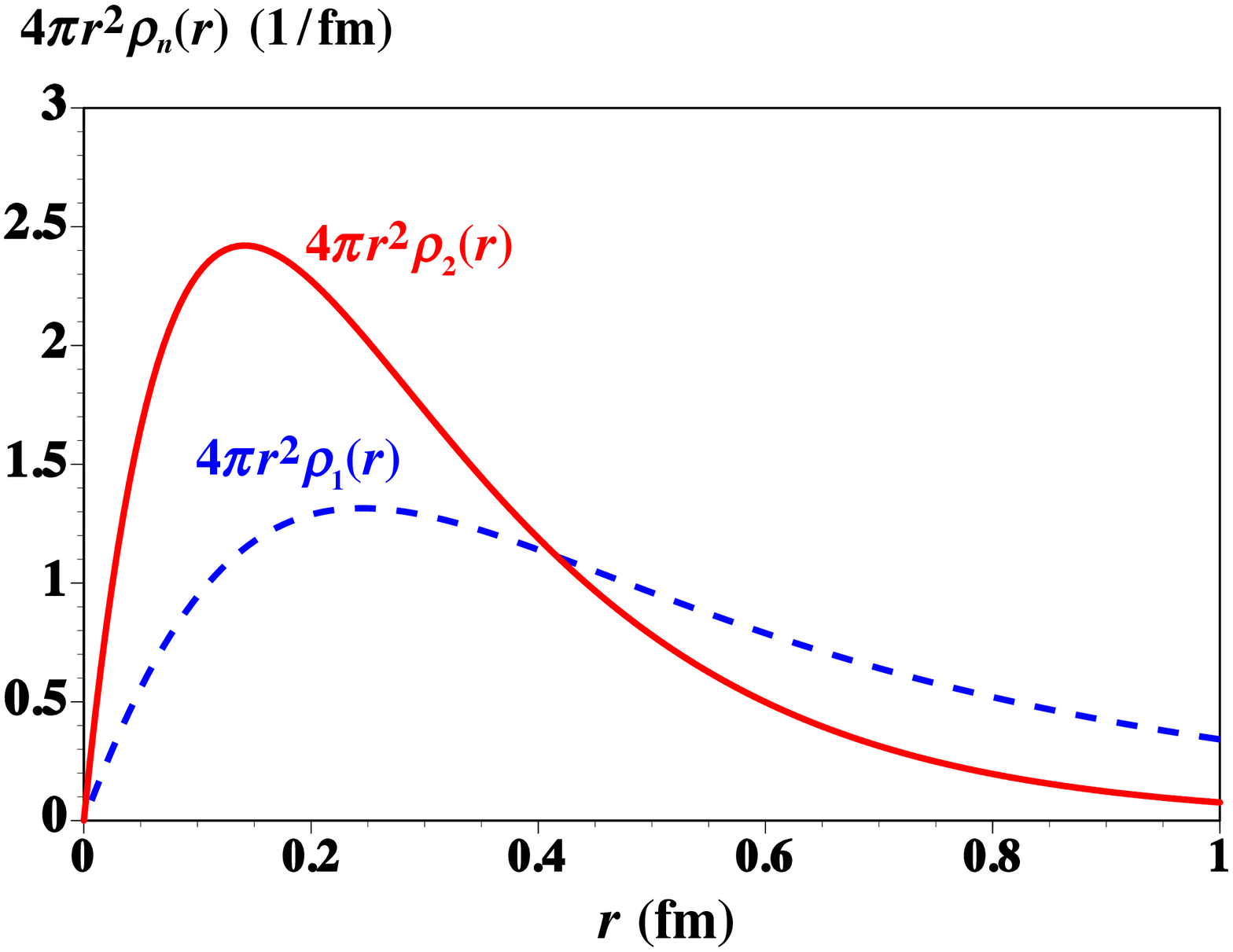}
\end{minipage}
\end{tabular}
\vspace{0.05cm}
\end{minipage}
\ \vspace{-0.00cm}
\hspace{3.75cm} $(a)$ \hspace{6.4cm} $(b)$
\vspace{-0.3cm}
\caption{$(a)$ Spacelike gravitational form factors \ \ 
 (b) Mass and mechanical densities \cite{gdas-kst-2017}.}
\label{fig:spacelike-fig}
\vspace{-0.0cm}
\end{figure}

Once the GPDs are obtained, we can calculate the timelike gravitational
form factors of the pion by using Eqs.\,(\ref{eqn:integral-over-z})
and (\ref{eqn:emt-ffs-timelike-0}). Our results are shown in 
Fig.\,\ref{fig:theta12reim-s}. Since it is a timelike process,
the form factors have imaginary parts. 
The form factor $\Theta_2$ comes from the D-wave term, and it is peaked
at the $f_2 (1270)$ resonance. On the other hand, 
the S-wave term also contributes to $\Theta_1$ 
in addition to the D-wave one,
so that the curves indicate complicated interference patterns.
The timelike gravitational form factors are converted to the spacelike
ones by using the dispersion relation, and the results are shown
in Fig.\,\ref{fig:spacelike-fig}$(a)$. Taking the Fourier transforms
of these functions, we obtain the space-coordinate densities
$\rho_1 (r)$ and $\rho_2 (r)$ for $\Theta_1$ and $\Theta_2$,
respectively, in Fig.\,\ref{fig:spacelike-fig}$(b)$.

In order to understand the physics meaning of the form factors
and their densities, we may define the static energy-momentum tensor as
$ T^{\mu\nu}_q (\vec r \,) = 
\int d^3 q / [(2\pi)^3 \, 2E] e^{i \vec q \cdot \vec r} 
\left\langle \pi^0 (p') \! \left| 
T^{\mu\nu}_q (0) \, \right| \! \pi^0 (p) \right\rangle $,
where $E=\sqrt{m_\pi^2 +\vec q^{\ 2}/4}$ \cite{static-form}.
The $\mu\nu = 00$ component satisfies the mass relation
$ \int d^3 r \, T^{00}_q (\vec r ) = m_\pi \Theta_{2,q} (0)$,
so that $\Theta_2$ and $\rho_2 (r)$ reflect
the mass (energy) distributions in the pion.
The $\mu\nu = ij$ ($i,\, j =1,\, 2,\, 3$) components are
expressed by the pressure $p(r)$ and shear force $s(r)$ as
$ T^{\, ij}_q (\vec r \,) = p_q (r) \, \delta_{ij} 
    + s_q (r) ( r_i r_j / r^2 - \delta_{ij} /3 )$,
which is expressed solely by $\Theta_1$.
Therefore, $\Theta_1$ and $\rho_1 (r)$ contain information on
pressure and shear-force distributions in the pion.

From the densities, the gravitational mass and mechanical (pressure, 
shear force) radii are calculated from 
$\rho_2 (r)$ (or $\Theta_2 (t)$) and $\rho_1 (r)$ (or $\Theta_1 (t)$).
We obtained
\begin{align}
\sqrt {\langle r^2 \rangle _{\text{mass}}} 
    =  0.56 \sim 0.69 \, \text{fm}, 
\ \ \ 
\sqrt {\langle r^2 \rangle _{\text{mech}}} 
   = 1.45 \sim 1.56 \, \text{fm} .
\\[-0.70cm]
\nonumber
\label{eqn:g-radii-pion-range}
\end{align}
They indicate that the mass radius is similar
or slightly smaller than the charge radius
$\sqrt {\langle r^2 \rangle _{\text{charge}}} =0.672 \pm 0.008$ fm
and that the mechanical radius is larger.
This is the first finding on gravitational radii from
actual experimental measurements.
In comparison with the charge radius, much details should be 
investigated further.

We believe that this kind of field has a bright prospect 
in understanding gravitational physics in terms of quarks and gluons
with experimental justifications.
The KEK-B factory has just upgraded, and the errors of 
Fig.\,\ref{fig:belle-comparison} should be significantly reduced
in the near future. Furthermore, other hadron-pair production
processes are being analyzed by the Belle collaboration, and
the future ILC project can probe different kinematical regions
of the GDAs, especially the continuum part.
The GDA studies had been purely theoretical projects
for a long time; however, time has come to investigate them 
in comparison with experimental measurements and to explore 
gravitational physics in the quark and gluon level.

\section{Summary}
\label{summary}
\vspace{-0.20cm}

We have determined the pion GDAs from the KEKB measurements on 
$\gamma^* \gamma \to \pi^0 \pi^0$. From the obtained GDAs, 
the timelike gravitational form factors $\Theta_1 (s)$ and $\Theta_2 (s)$
are obtained. Using the dispersion relation, we calculated spacelike
gravitational form factors $\Theta_1 (t)$ and $\Theta_2 (t)$
and spacial densities $\rho_1 (r)$ and $\rho_2 (r)$.
The functions $\Theta_2 (t)$ [$\rho_2 (r)$] and $\Theta_1 (t)$ [$\rho_1 (r)$]
indicate gravitational mass (energy) and mechanical 
(pressure, shear force) distributions, respectively. 
From these distributions, the gravitational radii were determined 
for the pion. They are 
$\sqrt {\langle r^2 \rangle _{\text{mass}}}  =  0.56 \sim 0.69 \ \text{fm}$
and
$\sqrt {\langle r^2 \rangle _{\text{mech}}}  = 1.45 \sim 1.56 \ \text{fm}$.
The mass radius is similar to the charge radius $0.672 \pm 0.008$ fm
or slightly smaller, and the mechanical radius is larger.
This kind of studies have bright prospect in creating the field of
gravitational physics in the elementary quark and gluon level.

\vspace{-0.10cm}
\section*{Acknowledgement}
\vspace{-0.20cm}
This work was supported by Japan Society for the Promotion of Science (JSPS)
Grants-in-Aid for Scientific Research (KAKENHI) Grant Number JP25105010.

\vspace{-0.10cm}


\end{document}